# New approach for SCR selection and optimization for Septum magnet power supply with high reliability

Ankit Vora#

#Shri Vaishnav Institute of Technology & Science, Indore, Madhya Pradesh, India

#ankit_vora@ieee.org

**Abstract** – A new approach for selection of Silicon-Controlled Rectifier (SCR) for switching high pulsed currents in the septum magnet power supply is described. In this approach, an attempt is made to select the SCR from its $I^2t$ rating given in data sheet. For this, a factor which we have called $I^2t$ derating factor is defined as the ratio of $I^2t$ rating of SCR to the $I^2t$ value of current pulse to be switched. Thus, the SCR to be used in power supply is selected from its $I^2t$ derating factor. Three different experiments were performed using SCR of different manufacturers, to estimate the value of $I^2t$ derating factors. In these experiments, the SCR was subjected to maximum thermal stress till SCR started showing degradation. The forward blocking voltage and the leakage current are the parameters used for characterizing the SCR before and after the stress. The results of these experiments are presented and discussed. This approach will be of great help when pulse rating of SCR, like, transient thermal impedance is not available. Finally, a method is presented which can contribute to design low-cost and high-reliability septum magnet power supply.

1. INTRODUCTION

The septum magnet power supplies are used in booster synchrotron and storage rings for injection and extraction of high energy electron bunches by deflecting them in the pulsed magnetic field. The injection or extraction takes place at the flat top of half sine wave which should be flat within 0.01% for the duration of passage of electron bunches. Therefore, in most of the septum magnet power supplies the duration of half sine wave is taken to be in the range 100 - 200 μs and the peak currents are in the range 5 - 10 kA for electron energy up to few GeV or even more than that. When SCR is used in such a situation, the dI/dt turn on dissipation is low and hence being not of much concern. In the situations where dI/dt turn on dissipation is high, e.g. in kicker supplies, a lot of work is done for selection of SCR and some of these works have been reported in the literature [l] - [3]. But in the situation when the dI/dt turn on dissipation is not significant and peak current is in the range of 5 – 10 kA and pulse repetition frequency (PRF) is low (1 - 10 Hz) then selection of proper SCR becomes a problem. For such low PRF, the RMS current becomes very low and if one selects the SCR according to RMS current rating, the size of SCR comes out to be very small which leads to a very high value of instantaneous current density in the device. At such high current density, the instantaneous junction temperature can rise to the destruction temperature of the device (~ 1400° C) and spoil it. Therefore, one has to select an SCR in which the current density is low enough such that the instantaneous junction temperature does not exceed the maximum allowable value and it can withstand the required peak current repeatedly. The instantaneous rise of junction temperature is estimated from the value of transient thermal impedance $Z_{\theta(t)}$ supplied by the manufacturers on demand. This value of $Z_{\theta(t)}$, is specified only for a pulse width of 10 ms and it may be entirely different for a pulse width of the order 200 μs. Moreover, when packaging of the silicon wafer is done by a secondary manufacturer (other than the principal manufacturer) the value of $Z_{\theta(t)}$, may be entirely different from the value specified by principal manufacturers. This may be the reason perhaps secondary manufacturers do not supply the value of $Z_{\theta(t)}$ even on demand. This problem of selecting SCR for switching high peak current at low PRF was discussed with many indigenous and principal manufacturers, but they could not provide any data for selecting the SCR for this particular application, and this led to the development of this work. However, some work has already been done in this direction by Mapham [4], [7] but it is not of much help in this particular situation where repetition rate is very low. Ikeda et al. [5] have reported the calculations of the current pulse rating of SCR from maximum allowable junction temperature using a combination of analytical and experimental methods. This approach may give accurate pulse rating for an SCR but by no means, it's straight forward to adopt as it requires some experimentation involving the measurement of junction temperature. A lot of work has been done for designing septum magnet power supply [8] – [14], but this is not at all related to high reliability with lost-cost design as septum magnets are core features of particle accelerators and hence its power supply. The present work provides an alternative approach for selecting an SCR from $I^2t$ rating given in data sheets. For this, we conducted experiments to optimize the size of SCR for switching high peak currents at low PRF, particularly, for application in septum magnet power supplies.

2. THE HYPOTHESIS

The $I^2t$ rating of an SCR specifies that the device can switch a half sinusoidal current of peak value $\sqrt{2}I$ for a time t and for a limited number of times in the course of operating life of the device. Keith H. Billings [6] has discussed this subject more elaborately. This rating is therefore called non-repetitive rating and is used to accommodate unusual circuit conditions such as over load protection, short circuits, etc. The manufacturers warn against the utilization of these rating for any repetitive applications. When maximum surge current is switched through the device, the instantaneous junction temperature of the device exceeds the maximum operating value (125° C) for a brief instant but remains below the temperature (~ 1000° C) at which SCR characteristics starts degrading [5]. However, if SCR switches a current pulse with an $I^2t$ value much less than the $I^2t$ rating of the device then instantaneous junction temperature will be much less than 1000° C, and the SCR can switch that value of peak current repetitively. From this, it can be concluded that if SCR switches a current pulse of $I^2t$ value which is some fraction of $I^2t$ rating of the device, then the SCR should survive indefinitely. Therefore, we introduce a factor which we call as $I^2t$ derating factor and is being defined as the ratio of $I^2t$ (SCR) to $I^2t$ (pulse). In order to determine this factor, we have conducted some experiments so that one can select the SCR from its $I^2t$ rating for switching high repetitive, pulsed current in the situation where turn on switching losses are insignificant and pulse repetition frequency is low (1 - 10 Hz). This factor signifies that higher its value, less degradation the SCR shows.

3. THE EXPERIMENT

In the experimental setup, the pulse width of half sine wave was 200 µs. In order to generate the half sinusoidal pulse, we have used the circuit shown in figure 1. A DC power supply charges a bank of energy storage capacitor C via a transistor switch which is turned off by the control circuit and isolated base drive when the capacitor is charged to the required voltage. When SCR ($D_1$) is triggered, the capacitor C and coil $L_2$ form a resonant circuit, and a half sinusoidal current passes through SCR. The $L_1$ and $D_2$ recover the reverse charge on the capacitor. The trigger unit generates a trigger pulse of rise time 50 ns and is connected to the gate of SCR through the transmission type pulse transformer. The pulse width of the trigger pulse is 20 µs and current can be varied from 100mA to 2A. The repetition rate can be varied from 1 pps to 10 pps. The peak current can be varied by varying the voltage on capacitor through the control unit, and it can be adjusted up to 5000A. A half sinusoidal current pulse of half width 200 µs and the peak value of 1000A has an $I^2t$ value of 100 $A^2s$. In order to determine the $I^2t$ derating factor, we collected inverter grade SCRs from different manufacturers with a value of average current ranging from 10A to 180A. The description of these SCRs is given in table 1. All the devices were characterized initially, and the characterization parameters were chosen to be forward blocking voltage, $I^2t$ rating of SCR, and the leakage current. There could be more characterization parameters, but these three parameters are the most sensitive indicators of degradation of an SCR. Moreover, they are easy to measure with fair accuracy. During the experiment, the wave forms of the voltage across SCR and the peak current were monitored continuously in order to ensure full conduction of SCR when triggered. With a view to determine a reasonable value of $I^2t$ derating factor, we conducted three different types of experiments.

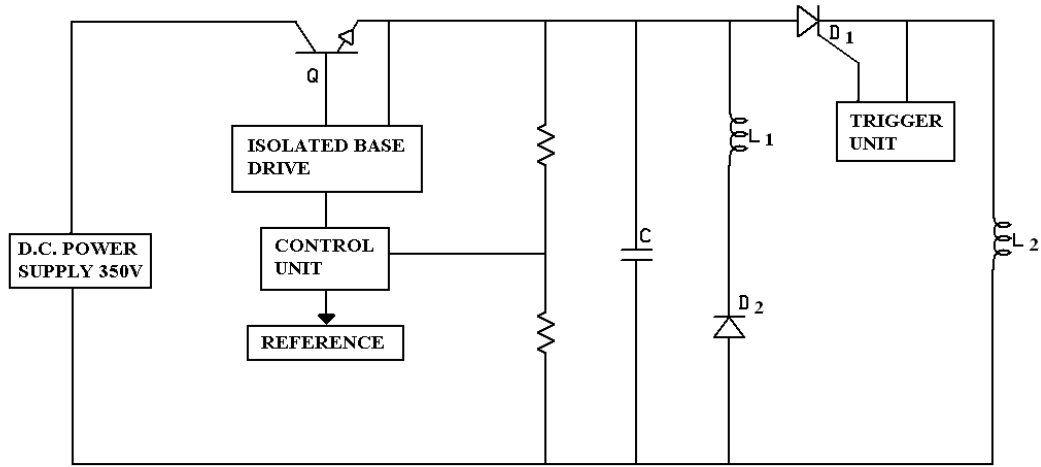

*Figure 1. Circuit diagram used for septum magnet power supply*

3.1 *Experiment – I*

*Table 1. Description of few selected SCRs*

| SCR TYPE | MANUFACTURER | AVERAGE CURRENT $I_{T(AV)}$ (A) | $I^2t$ ($A^2S$) |
|---|---|---|---|
| SKT 10 | Semikron | 14 | 310 |
| H16TB | Hind Rectifiers | 16 | 610 |
| 30TPS | Vishay | 20 | 442 |
| 22RIA | International rectifier | 22 | 560 |
| BHt E35 | BHEL | 25 | 910 |
| SKT 24 | Semikron | 29 | 1000 |
| SKT 40 | Semikron | 37 | 2500 |
| T500 | Powerex | 40 | 6000 |
| H45TBXX | Hind Rectifiers | 45 | 3200 |
| T510 | Powerex | 50 | 6000 |
| H55TBXX | Hind Rectifiers | 55 | 4050 |
| SKT 55 | Semikron | 65 | 8500 |
| BHt H05 | BHEL | 70 | 9520 |
| 80RIA | Vishay | 80 | 18000 |
| BHt H35 | BHEL | 85 | 13100 |
| C350PB | Powerex | 115 | 15600 |
| P0128SH10 | Westcode | 128 | 19000 |
| 180RKI | Vishay | 180 | 72000 |

All the available SCRs of average current ranging from 10A to 180A were characterized initially. These SCRs were used to switch a sinusoidal peak current of 1000 A and pulse width 200 µs, one by one. This experiment was performed at PRF of 1 Hz. The results of the experiment are summarized in Table 2. It was found that when $I^2t$ derating factor was below 2.55, the SCR failed catastrophically. When $I^2t$ derating factor was below 25, the SCRs did not fail catastrophically, but their forward blocking voltage dropped gradually as the number of shots increased. For the SCR type SKT 40 with $I^2t$ derating factor of 25, the leakage current increased 12-fold after $6 \times 10^4$ shots. The experiment was then repeated with another device (SKT 40) but peak current was set at 900A, the device survived $6 \times 10^4$ shots. The SCR with $I^2t$ derating factor above value 30 survived $10^7$ shots with no sign of degradation.

## 3.2 Experiment – II

In this experiment, only indigenous SCRs which survived indefinitely in experiment-I were used. The pulse repetition rate was kept at 1 Hz, and the starting value of peak current was set to be 800A for the device H45TBXX and 1400A for others. The $I^2t$ value of the pulse was increased by increasing the peak current in step of 100A, and the SCRs were re-characterized after $10^3$ shots. In this experiment, SCRs of same manufacturers were taken one by one and peak current was increased till the SCR started showing degradation. The degradation in forward blocking voltage was detected immediately during the experiment. When the forward blocking voltage dropped, the charging power supply was not able to charge the capacitor up to a value set by reference because the SCR conducted as soon as the capacitor voltage reaches the degraded value of the forward blocking voltage of SCR. From these observations, it was possible to determine a maximum value of peak current at which there was no degradation of any of characterizing parameters. From these values of peak current, the $I^2t$ derating factor was calculated for the used SCRs. The experiment was then repeated with other SCRs which survived in experiment-I.

## 3.3 Experiment – III

In this experiment, twenty new SCRs of each type and each manufacturer were used. The SCRs were mounted on a heat sink, and the temperature of the case was monitored. The pulse repetition rate was kept at 10 Hz. The peak current was set, at a value of which the SCR survived after $10^4$ shots for PRF of 1 Hz, from the readings of the experiment-II. The SCRs from each manufacturer were stressed under the above conditions. Half of the devices were operated for a total $10^6$ shots and stopped in between, for characterization after every $10^4$ shots. The second halves of the devices were operated for a total of $10^8$ shots and were characterized only at the end. Out of all survived devices in experiment I & II, one device H45TBXX showed a slight increase in leakage current. It means that set value of 900A for this device was perhaps critical at a repetition rate of 10 Hz. In all other devices, no degradation of any characteristic could be observed.

*Figure 2. SCRs from different manufacturers having same $I_{T(AVG.)}= 16A$ but different $I^2t$ rating, hence different $I^2t$ derating factors for them.*

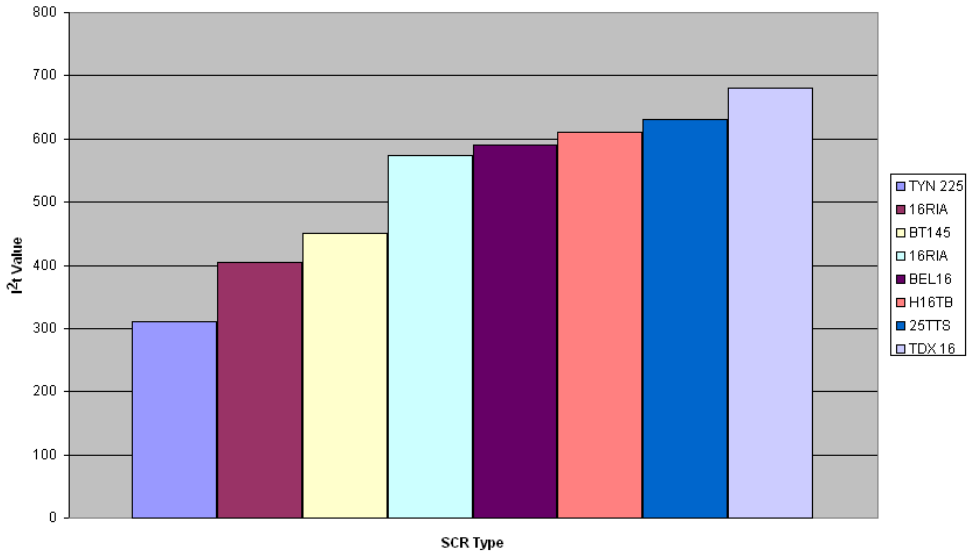

## 4. RESULTS AND DISCUSSION

The experiment-I was performed with devices of different $I^2t$ rating ranging from 180 $A^2s$ to 18000 $A^2s$ from various manufacturers. This experiment resulted in an approximate value for $I^2t$ derating factor $\geq 25$ (which survived). In the second experiment, the $I^2t$ value of pulse was increased till SCRs started degrading. Thus, we obtained fairly accurate value of $I^2t$ derating factor for each device with the number of shots limited to $10^7$ only (for survived devices). Obviously, these many shots are sufficient to conclude that $I^2t$ derating factor obtained in this condition is valid for repetitive application. To obtain even more accurate results, repetition rate was increased in the experiment-III. The SCRs were subjected to a total of approximately $10^8$ shots which is probably enough to conclude that $I^2t$ derating factors obtained are valid for repetitive applications. Not only this, the SCR is subjected to high thermal stress when it is switched at higher PRF. In this situation, the inter-pulse time becomes comparable with a thermal time constant of the path, junction-to-face of the silicon pellet and the heat generated during the pulse may not get sufficient time to transfer to the case, and junction temperature may be very close to maximum operating value. Therefore, it can be concluded from the survival of SCRs in the experiment- III that the values of $I^2t$ derating factors are repetitive in nature.

Based on these results, the SCR for injection septum magnet power supply was selected to be H45TBXX of make Hind Rectifier which is the result of proper selection and optimization as per the experiments conducted above. This power supply is designed to switch a peak current of 900A at a repetition rate of 1 pps and pulse width of half sinusoidal current pulse is 200 µs. In the extraction septum magnet power supply that we designed using this procedure, the maximum peak current which can be generated is 5000A at 1 pps. The SCR finally selected and employed in this power supply is of Hind Rectifier type H175TB with a specified $I^2t$ rating of 104x10$^3$ A$^2$s. The $I^2t$ value of pulse comes out to 2500 A$^2$s which corresponds to derating factor of approximately 40. The power supply is being used by injection septum magnet and has completed about $10^9$ shots to date. The SCR was re-characterized, and no degradation in any of characteristics could be noticed.

*Figure 3. Graph for $I^2t$ derating factor vs. different types of SCR used. Results show that any SCR can be used for making septum magnet power supply whose $I^2t$ derating factor is greater than 30. This ensures that it will survive indefinitely without any degradation.*

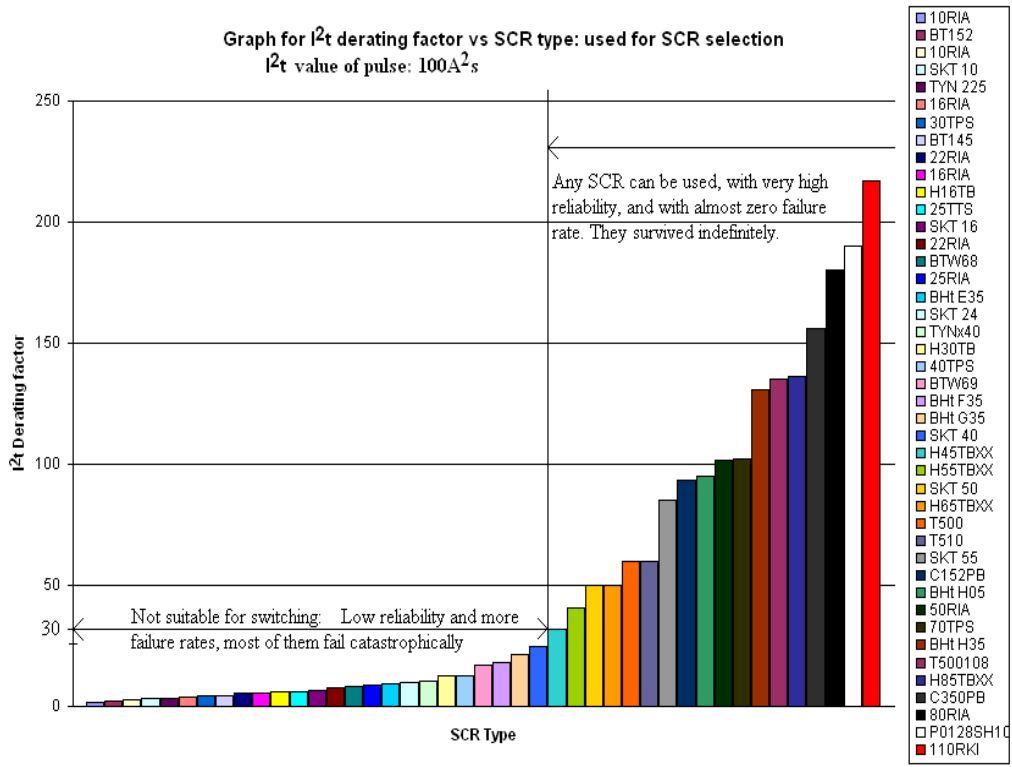

*Table 2. Results of Experiment I, II and III along with description of the few selected SCRs*

| SCR TYPE/ AVG. CURRENT $I_{T(AV)}$ (A) | | MANUFACTURER | $I^2t$ ($A^2S$) | $I^2t$ Derating Factor | Results |
|---|---|---|---|---|---|
| SKT 10 | 14 | Semikron | 310 | 3.10 | FBV dropped to 25 V after 180 shots |
| H16TB | 16 | Hind Rectifiers | 610 | 6.10 | FBV dropped to 36 V after $2.7 \times 10^3$ shots |
| 30TPS | 20 | Vishay | 442 | 4.42 | FBV dropped to 31 V after $1.6 \times 10^3$ shots |
| 22RIA | 22 | International rectifier | 560 | 5.60 | FBV dropped to 35 V after $2.9 \times 10^3$ shots |
| BHt E35 | 25 | BHEL | 910 | 9.10 | FBV dropped to 41 V after $4.8 \times 10^3$ shots |
| SKT 24 | 29 | Semikron | 1000 | 10.00 | FBV dropped to 43 V after $6.5 \times 10^3$ shots |
| SKT 40 | 37 | Semikron | 2500 | 25.00 | Survived but leakage current increased |
| T500 | 40 | Powerex | 6000 | 60.00 | Survived indefinitely |
| H45TBXX | 45 | Hind Rectifiers | 3200 | 32.00 | Survived indefinitely |
| T510 | 50 | Powerex | 6000 | 60.00 | Survived indefinitely |
| H55TBXX | 55 | Hind Rectifiers | 4050 | 40.50 | Survived indefinitely |
| SKT 55 | 65 | Semikron | 8500 | 85.00 | Survived indefinitely |
| BHt H05 | 70 | BHEL | 9520 | 95.20 | Survived indefinitely |
| 80RIA | 80 | Vishay | 18000 | 180.00 | Survived indefinitely |
| BHt H35 | 85 | BHEL | 13100 | 131.00 | Survived indefinitely |
| C350PB | 115 | Powerex | 15600 | 156.00 | Survived indefinitely |
| P0128SH10 | 128 | Westcode | 19000 | 190.00 | Survived indefinitely |
| 180RKI | 180 | Vishay | 72000 | 720.00 | Survived indefinitely |

## 5. CONCLUSION

In septum magnet power supplies where pulse width at the base of half sinusoidal pulse is in the order of few hundred microseconds, the turn-on switching dissipation is not dominating, and on-state conduction losses are the major source of junction heating. The SCR for maximum repetitive peak current can be selected from its $I^2t$ rating specified in manufacturers data sheet by suitable derating it. The derating factor for various SCRs was found to be in the range of 25 - 180. The $I^2t$ rating specified in the data sheet are derived from surge current rating, $I_{TSM}$ which is the maximum allowable non-repetitive value of half Sinusoidal current of duration 10 ms (half period of 50 Hz mains frequency). The $I_{TSM}$ is measured by monitoring the instantaneous rise in junction temperature which should be much less than destruction temperature for silicon. The safety margin in this instantaneous temperature may be different for different manufacturers. Another factor which may influence the measurement of $I^2t$ value may be the measurement of instantaneous junction temperature itself. The instantaneous junction temperature is measured by indirect methods such as by monitoring one of the temperature sensitive junction characteristic, which can be the on-state voltage drop across SCR. Therefore, the level for specifying the $I^2t$ rating may vary from one manufacturer to another for the same type of SCR. In figure 2, we have depicted the $I^2t$ ratings of SCR of same average current and nearly same steady state thermal impedance from junction to case assuming that transient thermal impedance will also be approximately same. It can be seen that $I^2t$ ratings vary a lot from one manufacturer to another. Therefore, for septum magnet power supplies or similar applications, the SCR should be selected which has a maximum $I^2t$ rating for given average current. Keeping in view these complexities, we have taken the more conservative value of $I^2t$ derating factor which comes out to be 30 or more for selecting the SCR for our septum magnet power supplies. In the case of higher PRF, of course, the r.m.s rating will also have to be considered. The approximate derating factor of 30 ensures that instantaneous temperature rise is lower than maximum safe junction temperature. Hence power supply with a high degree of reliability with very low cost can be constructed by selecting suitable SCR from its $I^2t$ derating factor.

ACKNOWLEDGMENTS

The authors wish to express sincere thanks to A.P. Thrpsay and A.C. Thakurta for helpful suggestions and discussions and Lal Viswanath for assistance in experiments.